\documentclass[twocolumn, a4paper, aps, prl, 10pt]{revtex4-1}

  \usepackage[T1]{fontenc}                     
  \usepackage{lmodern}                         
  \usepackage{graphicx}                        
  \usepackage{xcolor}                          
  \usepackage{amsmath}
  \usepackage{amsfonts}
  \usepackage{amssymb}
  \usepackage{hyperref} 
  \usepackage[]{siunitx} 

\newcommand{\abs}[1]{\ensuremath{ \left| #1 \right| }}                   
\newcommand{\mean}[1]{\ensuremath{ \left\langle#1\right\rangle }}         
\newcommand{\del}[0]{\partial}                
\newcommand{\eps}[0]{\varepsilon}
\renewcommand{\rho}[0]{\varrho}
\renewcommand{\theta}[0]{\vartheta}
\renewcommand{\phi}[0]{\varphi}

\renewcommand{\vec}[1]{\mathbf{#1}}        
 
\newcommand{\ba}[1]{\begin{align} #1 \end{align}}

\begin{document}

\title{Spin-wave localization in disordered magnets}

\author{Martin Evers$^1$, Cord A. M\"uller$^{1,2}$, and Ulrich Nowak$^1$}
\affiliation{$^1$Fachbereich Physik, Universit\"at Konstanz, 78457 Konstanz, Germany}%
\affiliation{$^2$Institut Non Lin\'eaire de Nice, CNRS and Universit\'e Nice--Sophia Antipolis, 06560 Valbonne, France}%

\date{\today}

\begin{abstract}
The effect of disorder on magnonic transport in low-dimensional magnetic materials is studied in the framework of a classical spin model. Numerical investigations give insight into scattering properties of the systems and show the existence of Anderson localization in 1D and weak localization in 2D, potentially affecting the functionality of magnonic devices. 
\end{abstract}

\maketitle

The propagation of spin waves \cite{Demokritov13_Magnonics} is in the focus of modern research because of its importance for spin caloritronic applications \cite{Bauer12_SpinCaloritronics, Uchida10_SpinSeebeckInsulatorTransversal, Schlickeiser14_DWMotionTempGradients, Jiang13_ImageingThermalDWMotion, Yan11_MagnonicDWMOtion} and for future information processing devices which might either rely on magnonic \cite{Serga10_YIG_Magnonics} instead of electronic transport or combine electronic with magnonic transport.
The dynamics of spin waves is mostly described by the Landau-Lifshitz-Gilbert equation, a nonlinear equation of motion that describes the wave propagation as well as some degree of dissipation, included phenomenologically either following Landau and Lifshitz \cite{Landau35_LL_equation} or Gilbert \cite{Gilbert55_Gilbert_equation}.
This dissipation limits the coherent wave propagation to a spatial scale set by the propagation length $\zeta$ that depends on the material properties, especially the damping constant \cite{Ritzmann14_MagnonPropagation, Hoffman13_TheoryLongSSE}.
The microscopic origin of the damping is inelastic scattering with, e.~g., phonons \cite{Ebert11_AbInitioGilbertDamping}. 
Static imperfections of the magnetic crystal, on the other hand,  induce elastic scattering, which has two effects. 
First, it turns ballistic into diffusive transport, and, second, it might suppress transport completely, as first shown by Anderson in 1958 for spin diffusion in disordered lattices \cite{Anderson58_AndLoc}.

Meanwhile it has been established  for many different kinds of waves that quenched disorder in combination with phase coherence can lead to a complete suppression of transport, confining eigenmodes to spatial regions 
of a finite extent given by the localization length $\xi_\text{loc}$ \cite{50yearsAnderson}.
In addition, there is also the weak-localization regime where diffusive transport still prevails, but mesoscopic effects of phase-coherent scattering can be observed \cite{Akkermans2007}. 
Arguably, the most famous phenomenon of that kind is coherent backscattering (CBS), an effect that relies on long-range phase coherence and can therefore be seen as a gauge of the microscopic processes that eventually entail Anderson localization. 
CBS produces an enhanced intensity for the elastic scattering of an excitation with wave vector $\vec{k}_0$ into the opposite direction $-\vec{k}_0$, and has been directly observed with, e.g., light \cite{Kuga1984,Albada1985,Wolf85_CBSPhotons,Gurioli2005}, acoustic \cite{Tourin1997,Weaver2000}, seismic \cite{Larose2004}, as well as matter waves \cite{Jendrzejewski2012a}.  
In contrast, localization phenomena for spin waves have been studied rather scarcely, mostly in amorphous materials with random anisotropy \cite{Bruinsma86_AndLocBreakdownOfHydrodynamics, Serota1988, Amaral1993} and by analogy with hard-core boson excitations on disordered lattices  \cite{Ma1986,Zhang1993,Yu2013,Alvarez2013}. 
But since different types of defects are very common in magnetic crystals, it is important to study their consequences for magnonic transport.
Indeed, a localization-induced breakdown of regular transport would severely hamper the functionality of real-world devices. 
It is the purpose of this paper to study localization phenomena with spin waves on the basis of numerical calculations. In particular it is important to determine the length scale of Anderson localization, and compare it with dissipative mechanisms which also limit magnonic transport \cite{Ritzmann14_MagnonPropagation, Hoffman13_TheoryLongSSE}.
Localization effects are known to be most relevant in low dimensional systems. We therefore investigate strong localization in one dimension and CBS in two dimensions. Also, experimental setups include often thin films or nanowires, which might be treated as low dimensional materials as soon as their extent is of the order of or lower than the wave length of the magnons.

\begin{figure*}[t]
  \includegraphics[width=\textwidth]{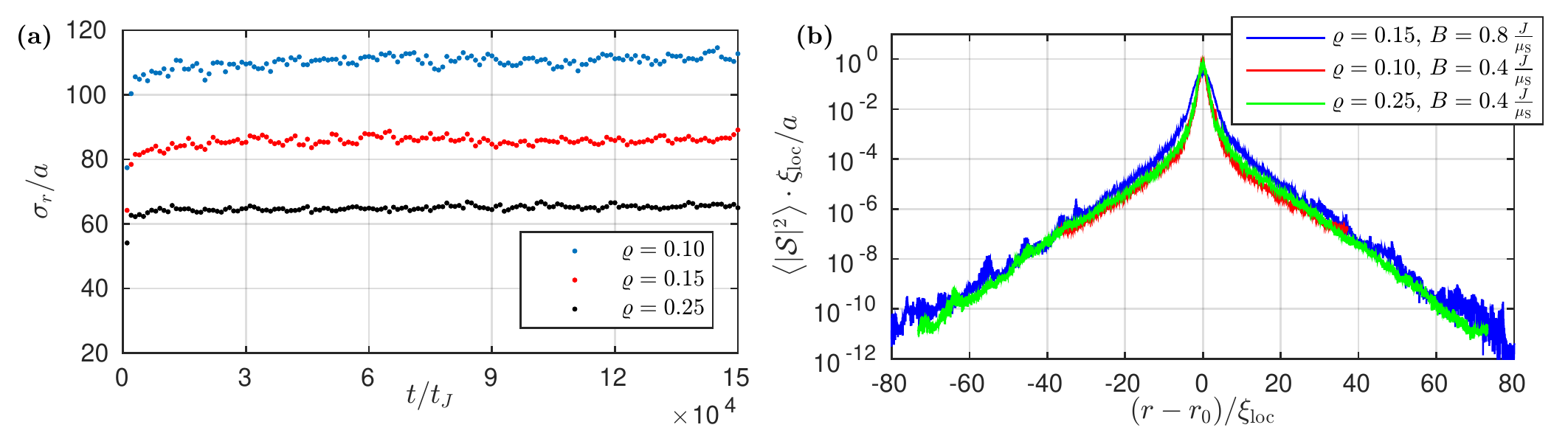}  
  \caption{%
  Strong localization of a spin wave packet, launched with initial wave vector $k_0=0.3/a$ (in units of lattice spacing $a$) and initial width $\sigma_0 = 50/\sqrt{2}\;a$ in a disordered 1D chain with a local field $B$ present on a fraction $\rho$ of random sites.
 (a) RMS wave packet spread $\sigma_r=[\langle{} r^2 \rangle{} - \mean{r}^2]^{1/2}$ as function of time for a defect field strength of $B=0.4\,J/\mu_\text{S}$. 
  At long times, magnonic transport comes to a halt. (b) Asymptotic in-plane magnetization profiles show exponential localization over the localization length $\xi_\text{loc}$.}
  \label{fig:1D_StrongLoc}
\end{figure*}

We study transport of magnons modeled as classical spin waves within the framework of an atomistic spin model \cite{Nowak07_SpinModels}.
The model comprises dimensionless magnetic moments (called ``spins'') $\vec{S}^l = \boldsymbol{\mu}^l/\mu_\text{S}$ on a lattice of sites $\vec{r}^l$, $l=1,\dots,N$,  with $\boldsymbol{\mu}^l$ the local magnetic moment and $\mu_\text{S}$ the reference value for the magnetic moment in the clean material. 
Each spin $\vec{S}^l$ has $N_\text{nb}$ neighbors $\vec{S}^{m}$, $m=1,...,N_\text{nb}$ at positions $\vec{a}^m$ relative to site $l$. 
The Hamiltonian of the system realizes a Heisenberg-type spin model 
\ba{
  H {=} \!\sum_{n=1}^N \!\bigg[\!{-}\frac{J}{2}\!\sum_{m=1}^{N_\text{nb}} \!\eps_n\eps_{m}\vec{S}^n{\cdot}\vec{S}^{m} {-} d_z \eps_n(S_z^n)^2 {-} \mu_\text{S} \eps_n\vec{B}^n{\cdot}\vec{S}^n\bigg]\!, \label{eq:Hamiltonian}
}
with ferromagnetic exchange interaction $J>0$, a uniaxial an\-iso\-tropy constant $d_z$ and an external magnetic field $\vec{B}^l$.
$\eps_l=0,1$ states the occupation of a site, which will be needed for defects.
The spins evolve in time according to a torque equation,
\ba{
  \frac{\del \vec{S}^l}{\del t} = - \frac{\gamma}{\mu_\text{S}} \vec{S}^l \times \vec{H}^l, \qquad \vec{H}^l = -\frac{\del H}{\del \vec{S}^l},
}
where $\gamma$ is the magnitude of the gyromagnetic ratio. This equation of motion corresponds to the Landau-Lifshitz (LL) equation in the limit of vanishing damping, where the total energy is conserved. 
The microscopic time scale of this model is $t_J = \mu_\text{S}/\gamma J$ ($\approx \SI{50}{\femto\s}$ for iron).
The natural order of magnitude for distances is the lattice constant $a\approx \SI{1}{\angstrom}$, and for local magnetic fields $J/\mu_\text{S}\approx \SI{100}{\tesla}$.

In the following we take the system to be globally magnetized along $z$ by choosing a small anisotropy $0 < d_z \ll J$.
Small-amplitude linear excitations, known as spin waves or magnons, are then confined to the $xy$-plane, $\mathcal{S}^l = S_x^l - iS_y^l$, such that the local wave intensity $|\mathcal{S}^l|^2$ measures the in-plane magnetization.
It will also prove fruitful to analyze the momentum-space  density $|\mathcal S_\vec{k}|^2$,  
where $\mathcal S_\vec{k}$ denotes the Fourier transform of $\mathcal S^l$. 
In the clean, simple cubic, $d$-dimensional lattice under consideration, the magnon dispersion 
reads 
\ba{
  \omega_\vec{k} = t_J^{-1} \left[ \frac{2d_z}{J} + \sum_{m=1}^{2d}\left(1-\cos\left(\vec{k}\cdot \vec{a}^m\right)\right) \right].  \label{omegak}
}
For infinitesimal anisotropy $d_z\ll J$, these spin waves are the gapless Goldstone modes of the ferromagnetic phase.  

The discrete translation symmetry of the lattice is broken by the presence of defects.  
We consider two kinds of uncorrelated defects, distributed with number density $\rho$ on randomly chosen sites $\vec{r}^j$.
One kind is a local magnetic field $\vec{B}^j = (0,0,B)$ along the easy axis, with $\vec{B}^l = 0$ everywhere else. (So here $\eps_l = 1\;\forall l$.)
The other kind is non-magnetic substitutional disorder with missing magnetic moments, $\eps_j = 0$ at defects sites and with $\eps_l = 1$ everywhere else. 
Since we investigate finite systems, calculated quantities usually depend on the defect configuration, 
and all results presented below will include an ensemble average, noted $\mean{...}$. 

We study the out-of-equilibrium, long-time spin wave dynamics, by integrating the equations of motion \eqref{eq:Hamiltonian} numerically, using an implicit Adams scheme.
The initial condition at time $t=0$ is a quasi-mono\-chro\-mat\-ic Gaussian wave packet 
\ba{
  \mathcal{S}^l_0 = A 
   \exp\big[i\vec{k}_0\cdot\vec{r}^l - (\vec{r}^l-\vec{r}_0)^2/4\sigma_0^2\big]
  \label{eq:InitCond}
}
with amplitude $A$ (in the range $0.01 \dots 0.1$) and rms width $\sigma_0$ around the initial position $\vec{r}_0$,   
launched with finite wave vector $\vec{k}_0$ into the bulk disordered lattice.

In a first step, we study a one dimensional spin chain, where disorder should manifest as strong localization.
We model defects by a random local field along $z$, taking a finite value $B$ with probability $\rho$ and vanishing with the complementary probability $1-\rho$. (While the other defect model of local missing spins is arguably more realistic in the bulk, we do not consider it for the 1D case since it breaks the exchange coupling and thus trivially confines the excitations to disconnected segments.) 
Our 1D simulation results are summarized in Fig.~\ref{fig:1D_StrongLoc}. Initially, the rms width of the wave-packet spreads in time, as shown in panel a). At longer times, the width saturates, and the spreading comes to a complete halt, which is a hallmark of localization.
As a rule, the higher the defect density, the stronger the localization effect, and the smaller the final extent.
The localization scenario is further corroborated by the asymptotic in-plane magnetization profiles, plotted in panel b) for different values of defect concentration and defect strength on a log-linear scale.
All profiles show the same characteristic exponential decrease, and a fit to the expected asymptotic form $\exp(-|r-r_0|/4\xi_\text{loc})$ \cite{Beenakker1997,Gogolin76_ElectronDensityLocalizedStates1D,Izrailev97_1DLocalizedWavePaket_Evoloution} yields the localization length $\xi_\text{loc}$.
The conventional factor 4 here emphasizes that the ensemble-averaged intensity decays more slowly than the typical (i.e.~most probable) intensity, which is approximately log-normal distributed and decreases as $\exp(-|r-r_0|/\xi_\text{loc})$ \cite{Beenakker1997,Mueller11_LocalizationPhenomena}.
Note that the localization lengths found in our simulations are at the order of $10^2$--$10^3\,a$ ($\approx\num{0.01}\text{--}\SI{0.1}{\micro\meter}$) for the chosen parameters and are therefore far below magnon propagation lengths $\zeta$ limited by dissipation, which can be in the range of $10^4\,a$ ($\approx\SI{1}{\micro\meter}$) for small damping constants $\alpha = 10^{-3} \dots 10^{-4}$ \cite{Ritzmann14_MagnonPropagation}.

The height of the wings depends on the distribution details near the center, which are found to deviate from the simple exponential cusp predicted in Ref.~\cite{Gogolin76_ElectronDensityLocalizedStates1D}. This is not surprising, given that our simulations  do not match the assumptions of the analytical calculations.
Notably, the wave packet starts with finite initial velocity and covers a certain range of momenta and energies. Moreover, the disorder parameters situate the simulation far from the perturbative regime. 
In particular, $\xi_\text{loc}$ cannot be expected to be given by the lowest-order term of an expansion in the defect strength of independent scatterers. 
Interestingly, deviations from the profile predicted by 
Ref.~\cite{Gogolin76_ElectronDensityLocalizedStates1D} have also been observed in numerical simulations of matter waves in uncorrelated on-site disorder \cite{Lee14_LocalizedWaves1D_CBS_CFS}.  In any case, the strong disorder prevents the system from reaching its equilibrium configuration (all spins aligned along $z$).
Instead, an in-plane magnetization remains forever written into the spin chain, thus highlighting the lack of ergodicity, one of the chief manifestations of localization \cite{Basko2006,Pal2010,Luca2013,Kjall2014}

\begin{figure}[t]
  \includegraphics[width=0.480\textwidth]{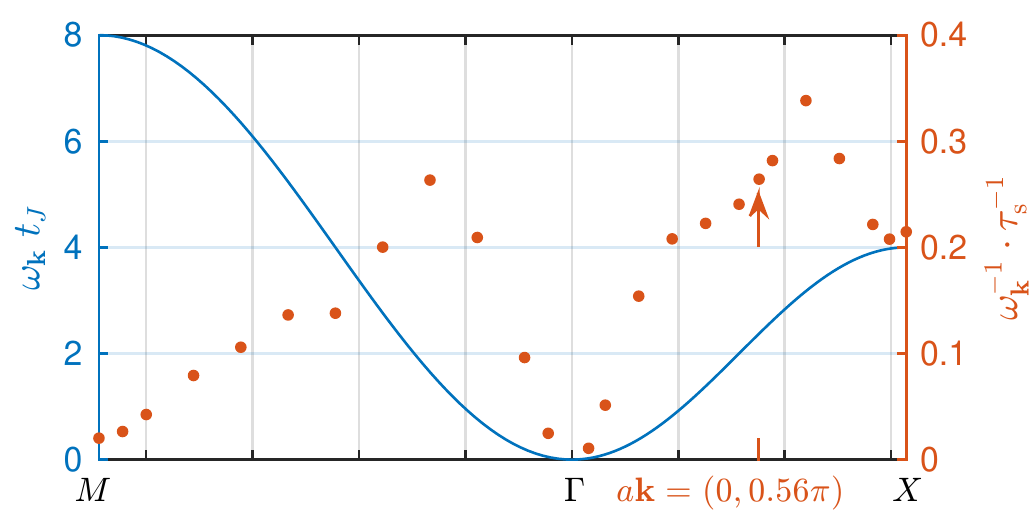}
  \caption{Clean dispersion relation $\omega_\vec{k}$, Eq.~\eqref{omegak}, and elastic scattering rate $1/\omega_\vec{k}\tau_\text{s}$  for a particular set of wave vectors inside the first Brillouin zone of a simple cubic lattice. Data points are obtained for a missing-spin defect density of $\rho=0.1$ and are averaged over 20 disorder configurations. 
  The scattering rate $\tau_\text{s}^{-1}$ vanishes at the central $\Gamma$-point. Strong scattering occurs for intermediate wave vectors.}
  \label{fig:ScatteringTime}
\end{figure}

In a second step, we analyze magnon scattering in a 2D disordered lattice with the aim of assessing weak localization effects. 
Since in magnetic materials non-magnetic defects are rather common, we place zero spins on randomly chosen lattice sites.
The only parameter describing the degree of disorder is therefore the defect density, or percentage of defect sites, which we take to be $\rho=0.1$ for the data presented below.
In order to gain a better understanding of the microscopic scattering processes at work for this type of disorder, we study the magnon dynamics in $\vec{k}$-space by evaluating $I_\vec{k}(t) = \mean{|\mathcal{S}_\vec{k}(t)|^2}$.
The initial wave packet, Eq.~\eqref{eq:InitCond}, is a peak of width $\sigma_0^{-1}$ centered at $\vec{k}_0$.
Due to elastic scattering off the defects, the initial wave packet is depleted, and the peak height decreases as $\exp(-t/\tau_\text{s})$, where $\tau_\text{s}$ is the elastic scattering mean free time.
This characteristic time can be measured by a fit to the observed exponential decay, thus revealing whether scattering can be qualified weak ($\omega_{\vec{k}}\tau_\text{s}\gg 1 $) or must be considered strong ($\omega_{\vec{k}}\tau_\text{s} \sim 1$). 

Fig.~\ref{fig:ScatteringTime} plots the dispersion, Eq.~\eqref{omegak}, together with the reduced scattering rate $1/\omega_{\vec{k}}\tau_\text{s}$ for selected 
wave vectors in the first Brillouin zone.
Near the symmetry point $\Gamma$ in the band center, the scattering amplitude from $\vec{k}$ to $\vec{k}'$ is proportional to $\vec{k}\cdot\vec{k}'$, as characteristic of p-wave scattering.
According to Fermi's Golden Rule \cite{Akkermans2007}, considering that the density of states in 2D is constant at low energy, one should expect the scattering rate to vanish like $|\vec{k}|^4$.
And indeed, a quadratic behavior of $1/\omega_\vec{k}\tau_\text{s}$ around the origin is consistent with the data.  
For comparison, we also determined the scattering rate for the local magnetic field defects, which produce an isotropic, s-wave scattering amplitude.
Consequently, the scattering rate decreases more slowly with $|\vec{k}|$, but vanishes nonetheless.
At first sight, this is at odds with Ref.~\cite{Bruinsma86_AndLocBreakdownOfHydrodynamics} when translated from 3D to our 2D case with its constant density of states.
But, as $\vec{k}\to0$, the independent-scatterer approximation used in Ref.~\cite{Bruinsma86_AndLocBreakdownOfHydrodynamics} breaks down, and collective scattering from impurity clusters eventually leads to a vanishing scattering rate, as expected from general principles for Goldstone modes at low energy \cite{Gurarie2003}. 
The local maxima of the scattering rate roughly halfway through the band, signaling strong scattering, may be traced back to 
a van Hove divergence of the density of states at frequency $\omega = 4\,t_J^{-1}$, shifted down in energy and rounded by the disorder. 
 
\begin{figure*}
  \includegraphics[width=\textwidth]{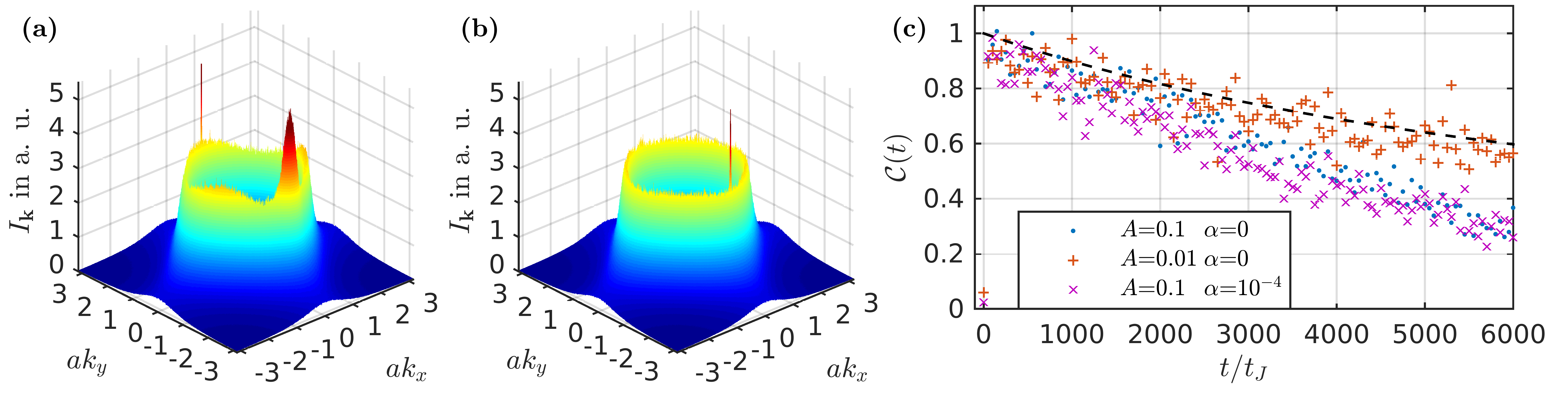}
  \caption{Spin-wave intensity $I_\vec{k} = \mean{\left|\mathcal{S}_\vec{k}(t)\right|^2}$ in 2D $\vec{k}$-space (average over 800 defect configurations) at times (a) $t=10\,t_J$ and (b) $t=1000\,t_J$. The initial wave packet can still be seen in (a) as the narrow peak at $a\vec{k}_0 = (0,0.56\pi)$.
  The CBS peak at $-\vec{k}_0$ is well visible above the diffusive background, distributed along the energy shell $\omega_\vec{k}= \omega_{\vec{k}_0}$. The width and contrast $\mathcal{C}$ of the CBS peak decrease in time. (c) Time evolution of the contrast for different settings, evaluating the impact of a weak nonlinearity ($A=0.1$) and finite damping ($\alpha>0$). 
The dashed line shows the diffusive prediction, Eq.~\eqref{eq:CBS_contrast}, for the linear case. 
Whereas the damping does not affect the CBS contrast, nonlinearities induce dephasing and reduce the CBS contrast noticeably. 
}
  \label{fig:CBS_2D}
\end{figure*}

For further investigation of the strong-scattering re\-gime, we take $\sigma_0 = 150/\sqrt{2}\,a$ and $a\vec{k}_0 = (0,0.56\pi)$ where $\tau_\text{s} \approx \num{1.59}\, t_J$ (see highlighted point in Fig.~\ref{fig:ScatteringTime}),
such that the elastic scattering mean-free path $l_\text{s}  = \abs{\vec{v}_\vec{k}} \tau_\text{s}\approx \num{3.1}\, a$ ($\vec{v}_\vec{k} = \partial_\vec{k} \omega_\vec{k}$ is the group velocity). 
Due to multiple elastic scattering, partial wave amplitudes appear in modes $\vec{k}$ with the same frequency $\omega_\vec{k} = \omega_{\vec{k}_0}$, up to a disorder-broadening of order $\tau_s^{-1}$.
Thus, we can follow the progressive, diffusive redistribution of wave vectors over the energy shell.
For a classical random-walk model, i.e., phase-incoherent propagation, one would expect a homogeneous distribution over all accessible modes, as a consequence of ergodicity.
A failure of ergodicity, instead, should show up as distinctive features in the wave-vector distribution $I_\vec{k}$. 

Fig.~\ref{fig:CBS_2D}(a) shows $I_\vec{k}(t)$ at a time $t=10\,t_J\gg\tau_\text{s}$ well in the diffusive regime.
Above the diffusive background of height $I_\text{bg}$, it clearly features the CBS peak at $-\vec{k}_0$, whose presence proves that the memory of the initial condition is preserved for very long times. 
Starting from a pure plane-wave excitation, the peak contrast $\mathcal{C}=(I_{-\vec{k}_0}-I_\text{bg})/I_\text{bg}$ with respect to the background should be exactly unity and constant in time. 
This signal has to be convolved with the initial $\vec{k}$-space distribution following from Eq.~\eqref{eq:InitCond}, and therefore is expected to decrease as  \cite{Cherroret12_CBS_MatterWavesTheory}
\ba{
  \mathcal{C}(t) = 
  \frac{4\sigma_0^2}{3\sigma_0^2 + \sigma^2(t)}, 
  \label{eq:CBS_contrast}
}
where the diffusive spread $\sigma^2(t) =\sigma_0^2+\abs{\vec{v}_\vec{k}}^2\tau_\text{tr}t$ increases linearly in time.
In the case investigated here, we determine the transport time $\tau_\text{tr}$ using a Green-Kubo relation
\ba{
  \mean{\vec{v}_\vec{k} \cdot \vec{v}_{\vec{k}_0}}(t) \!:=\! \frac{1}{\mathcal{N}}\!\!\sum_{\vec{k}\in\text{B.Z.}} \!\!\!I_\vec{k}(t)\,\vec{v}_\vec{k} \cdot \vec{v}_{\vec{k}_0} \propto \exp(-t/\tau_\text{tr}),
}
where $\mathcal{N}=\sum_\vec{k} I_\vec{k}$ is the (time-independent) normalization. For the chosen parameters we obtain $\tau_\text{tr}\approx 1.3\,t_J$.
Fig.~\ref{fig:CBS_2D}(c) shows the observed contrast, together with the prediction \eqref{eq:CBS_contrast}, for different simulation parameters.
The agreement is excellent for a small spin wave amplitude $A=0.01$. 
For a larger amplitude $A=0.1$ the CBS contrast decays faster, a behavior that we attribute to the dephasing caused by the nonlinearity (such effects are known for, e.g., light \cite{Chaneliere2004} and matter waves \cite{Hartung2008}). 

The third set of parameters includes finite damping.
Indeed, more realistic dynamic models for spin waves include the Gilbert damping via an additional term in the Landau-Lifshitz-Gilbert (LLG) equation, proportional to the damping constant $\alpha$ \cite{Nowak07_SpinModels}.
The observed contrast of the CBS peak, for the chosen value of $\alpha=\num{1e-4}$, remains unchanged compared to the undamped case.
This is in agreement with the reciprocity principle, well known in optics \cite{vanTiggelen1998_Reciprocity}, namely that uniform damping lowers the overall intensity, but preserves the CBS contrast compared to the background.

In summary, we have numerically studied the influence of random defects on the propagation of classical spin waves in low-dimensional disordered magnets.
We find evidence for strong (Anderson) localization of spin waves in a one dimensional spin chain.
In a two dimensional disordered lattice, a clear coherent backscattering signal proves the presence of weak localization effects---a well-known precursor for Anderson localization.
These findings underpin the importance of defect-related effects on magnonic transport, and define new limits for the propagation of spin waves in addition to the usually assumed Gilbert damping. 
In dimensions higher than one, the crossover to the strongly localized regime is hard to reach by direct numerical integration because it typically occurs at much longer times and for much larger system sizes. 
If one tries to increase the fraction $\rho$ of missing spins too much, the lattice becomes disconnected at the percolation threshold, and the excitations become trivially confined to the percolation clusters, which was not the regime of interest here. 
In order to compute localization lengths in the linearized regime (together with critical properties of possible localization-delocalization transitions in higher dimensions), the method of choice is a transfer-matrix 
approach combined with a finite-size scaling analysis \cite{MacKinnon1983,Slevin2014}. Nonlinearities, on the other hand, generically suppress the onset of Anderson localization, and lead to subdiffusive behavior instead \cite{Cherroret2014}. The quantitative investigation of such effects in substitutionally disordered magnets poses interesting challenges for future work.

\acknowledgments{%
This work was performed on the computational resource bwUniCluster funded by the Ministry of Science, Research and Arts and the Universities of the State of Baden-W\"urttemberg, Germany, within the framework program bwHPC.
Financial support by Deutsche Forschungsgemeinschaft (DFG) via SPP 1538 ``Spin Caloric Transport'' and SFB 767 ``Controlled Nanosystems: Interaction and Interfacing to the Macroscale.'' is gratefully acknowledged. 
}

\bibliography{../Bibliography}

\end{document}